\documentclass[aps,prl,reprint,amsmath,amssymb,amsfonts,showkeys,showpacs,keywords]{revtex4-1}
\usepackage{graphicx,dcolumn,bm,color,hyperref,microtype,multirow,subfigure}

\begin{document}
\newcommand{\EHF}{E^{\rm HF}}
\newcommand{\Ec}{E^{\rm corr}}
\newcommand{\Eh}{E_{\rm h}}
\newcommand{\rij}{\hat{r}}
\newcommand{\mc}{\multicolumn}
\newcommand{\mr}{\multirow}
\newcommand{\alert}[1]{\textcolor{black}{#1}}

\title{Exact wave functions of two-electron quantum rings}

\author{Pierre-Fran\c{c}ois Loos}
\email{loos@rsc.anu.edu.au}
\author{Peter M.W. Gill}
\thanks{Corresponding author}
\email{peter.gill@anu.edu.au}
\affiliation{Research School of Chemistry, Australian National University, Canberra ACT 0200, Australia}
 
\date{\today}

\begin{abstract}
We demonstrate that the Schr\"odinger equation for two electrons on a ring, which is the usual paradigm to model quantum rings, is solvable in closed form for particular values of the radius. We show that both polynomial and irrational solutions can be found for any value of the angular momentum and that the singlet and triplet manifolds, which are degenerate, have distinct geometric phases. We also study the nodal structure associated with these two-electron states.
\end{abstract}

\maketitle

%%% INTRODUCTION %%%
{\em Introduction.---}Like quantum dots \cite{Reimann02}, quantum rings (QR) are self-organized nanometric semiconductors, and are intensively studied experimentally due to their rich electronic, magnetic and optical properties \cite{Warburton00, Lorke00, Fuhrer01, Bayer03, Fuhrer04, Sigrist04}, such as the Aharonov-Bohm effect \cite{Aharonov59, Aronov93, Morpurgo98}.

Many-electron QRs have been investigated theoretically using various methods, such as model Hamiltonian \cite{Viefers04, Fogler05, Fogler06}, exact diagonalization \cite{Niemela96, Gylfadottir06}, quantum Monte Carlo \cite{Gylfadottir06, Emperador03}, and density-functional theory \cite{Emperador01, Rasanen09, Aichinger06, Manninen09} (DFT). Accurate numerical calculations on two-electron QRs have been reported in Ref.~\cite{Zhu03}.

Quantum rings are usually modelled by electrons confined to a strict- or quasi-one-dimensional circular space interacting via a short-ranged or Coulomb operator. In this Letter, we focus on the simple system in which two electrons are confined to a ring of radius $R$ and interact via a Coulomb operator. This choice has often been avoided in the literature due to the divergence of the Coulomb interaction at small interelectronic distances.

Contrary to frequent claims, systems with two electrons do not inevitably have intractable Schr\"odinger equations and we show here that, for each electronic state of a two-electron QR, the Schr\"odinger equation can be solved exactly for a countably infinite set of $R$ values, yielding both polynomial and irrational solutions in terms of the interelectronic distance.  Quantum mechanical systems whose Schr\"odinger equations can be solved in this way, such as the Hooke's law \cite{Taut93} or spherium \cite{QuasiExact09, Excited10} atoms, have ongoing value both for illuminating more complicated systems \cite{EcLimit09, EcProof10} and for testing and developing theoretical approaches, such as DFT \cite{Parrbook, Filippi94, Glomium11, UEGs12} and explicitly correlated methods \cite{Kutzelnigg85}.

In atomic units ($\hbar=m=e=1$), the Hamiltonian of two electrons on a ring of radius $R$ is
\begin{equation}
	\label{H}
	\hat{H} = \frac{1}{2} \left(p_1^2 + p_2^2\right)+ \frac{1}{u},
\end{equation}
where $p_k = (i/R) \partial/\partial\theta_k$ is the momentum operator associated with the electron $k$, and $\theta_k$ is its angle around the ring center. The operator $u^{-1}$ represents the Coulomb interaction between the electrons, where
\begin{equation}
	\label{u}
	u = R \sqrt{2-2\cos\left(\theta_1 - \theta_2\right)}
\end{equation}
is the interelectronic distance  \footnote{\alert{Spin-orbit effects can be taken into account using the Rashba coupling.}}. In one dimension, the singlet and triplet manifolds are degenerate \footnote{In one dimension, the divergence of the kinetic energy is unable to compensate the divergence of the Coulomb potential. Thus, the wave function has a node at $u=0$, the singlet (S) and triplet (T) states are degenerate \cite{Lee11}, and we have $\Phi_{\text{T}} = \text{sgn}(\theta_1 - \theta_2) \Phi_{\text{S}}$.} and this allows us to focus primarily on the singlets.

%%% HARTREE-FOCK SOLUTION %%%
{\em Hartree-Fock solution.---}Within the Hartree-Fock approximation \cite{SzaboBook}, the ground-state wave function is simply
\begin{equation}
	\label{Phi-HF}
	\Psi_{\text{HF}}(u) = u,
\end{equation}
which has a node at $u=0$, and the energy is
\begin{equation}
	\epsilon_{\text{HF}} =  \frac{1}{4R^2} + \frac{2}{\pi R}.
\end{equation}

%%% TABLE I %%%
\begin{table}
\caption{
\label{tab:states}
Term symbols for two electrons on a ring.  Geometric phases are indicated by the sign at the bottom-left of each symbol.  A $-$ sign means that the wave function changes sign when one of the electrons rotates once around the ring.}
\begin{ruledtabular}
\begin{tabular}{ccccccc}
Spin		&									\mc{6}{c}{Angular momentum $J$}											\\
\cline{2-7}
manifold	&			0		&		1		&			2		&		3			&			4			&	\ldots	\\
\hline
Singlet		&	$_+^1\Sigma$	&	$_-^1\Pi$	&	$_+^1\Delta$	&	$_-^1\Phi$		&	$_+^1\Gamma$	&	\ldots	\\
Triplet		&	$_-^3\Sigma$	&	$_+^3\Pi$	&	$_-^3\Delta$	&	$_+^3\Phi$		&	$_-^3\Gamma$		&	\ldots	\\
\end{tabular}
\end{ruledtabular}
\end{table}

%%% EXACT SOLUTION %%%
{\em Exact solution.---}In terms of the extracule coordinate $\Omega = (\theta_1 + \theta_2)/2$ and intracule coordinate $\omega = \theta_1 - \theta_2$ \alert{\cite{Leinaas77}}, the Hamiltonian \eqref{H} is $\hat{H} = \hat{H}_{\Omega} + \hat{H}_{\omega}$, where
\begin{align}
	\label{H-Omega}
	\hat{H}_{\Omega} & = - \frac{1}{4R^2}\frac{\partial^2}{\partial \Omega^2},
	\\
	\label{H-omega}
	\hat{H}_{\omega} & =  - \frac{1}{R^2}\frac{\partial^2}{\partial \omega^2},+ \frac{1}{R\sqrt{2-2\cos\omega}}.
\end{align}
The exact spatial wave function is then the product
\begin{equation}
\label{Phi}
	\Phi(\Omega,\omega) = \Lambda(\Omega) \Psi(\omega),
\end{equation}
and the exact total energy is the sum $E = \mathcal{E} + \epsilon$ of the extracular and intracular energies.

The eigenfunctions and eigenvalues of $\hat{H}_{\Omega}$ are
\begin{align}
	\Lambda_J(\Omega) & = \exp\left(i J\Omega\right),
	& 
	\mathcal{E}_J & = \frac{J^2}{4R^2},
\end{align}
where $J \in \mathbb{N}$ is the total angular momentum associated with the center-of-mass coordinate \footnote{A magnetic field of magnitude $B$ perpendicular to the ring can be applied to the system. It only influences the angular part of the Hamiltonian $\hat{H}_{\Omega}$, and has the effect of shifting the angular momentum $J$ by a value proportional to $B$. See Ref.~ \cite{Zhu03} for more details. \alert{Alternatively, one can introduce the Aharonov-Bohm phase using a unitary transformation \cite{Viefers04}.}}.

The eigenfunctions of $H_\omega$ satisfy
\begin{equation}
\label{H-intracule}
	-\frac{\Psi^{\prime\prime}(\omega)}{R^2}
	+ \frac{\Psi(\omega)}{R\sqrt{2-2\cos\omega}} = \epsilon\,\Psi(\omega).
\end{equation}
and are all doubly degenerate.  Each pair of solutions consists of a singlet ($S=0$) and a triplet ($S=1$) state with opposite Berry \cite{Berry84} (or geometric \cite{PhaseBook}) phase behavior (Table \ref{tab:states}).  Specifically, if one of the electrons passes once around the ring, the wave function is unaffected in states where $J+S$ is even but changes sign in states where $J+S$ is odd.  Two-electron QRs are probably one of the simplest systems that exhibit the Berry phase phenomenon.

In terms of $u$, Eq.~\eqref{H-intracule} is the Heun-like equation \cite{Ronveaux, NISTbook}
\begin{equation}
\label{H-u}
	\left(\frac{u^2}{4R^2}-1\right) \Psi^{\prime\prime}(u)  + \frac{u}{4R^2} \Psi^{\prime}(u) + \frac{\Psi(u)}{u} =  \epsilon\,\Psi(u),
\end{equation}
with singular points at $-2R$, $0$ and $2R$. A Kato-like analysis \cite{Kato57} of \eqref{H-u} yields 
\begin{align}
	\Psi(0) & = 0,	
	&
	\frac{\Psi^{\prime\prime}(0)}{\Psi^{\prime}(0)} & = 1,
\end{align}
which shows that, like $\Psi_{\text{HF}}$, $\Psi$ has a node at $u=0$, and behaves as 
\begin{equation}
\label{Kato}
	\Psi(u) = u\left(1+\frac{u}{2}\right) +O(u^3)
\end{equation}
for small $u$.

The general solution of \eqref{H-u} is \cite{NISTbook} 
\begin{equation}
\label{Psi-series}
	\Psi^{(a,b)}(u) = u\left(1+\frac{u}{2R}\right)^{a/2}\left(1-\frac{u}{2R}\right)^{b/2} P^{(a,b)}(u),
\end{equation}
where $a,b=0$ or $1$, and $P^{(a,b)}$ is a regular power series
\begin{equation}
\label{Pu}
	P^{(a,b)}(u) = \sum_{k=0}^{\infty} c_{k}^{(a,b)}\,u^k.
\end{equation}
This produces four families of solutions characterized by the ordered pair $(a,b)$. Substitution of \eqref{Psi-series} into \eqref{H-u} yields the three-term recurrence relation
\begin{multline}
\label{recurrence}
	c_{k+2}^{(a,b)} = \frac{1}{(k+2)(k+3)}\left\{\left[\frac{(k+2)(b-a)}{2R}+1\right] c_{k+1}^{(a,b)} \right.
	\\
	+ \left. \left[\frac{k(k+2+a+b)+\sigma^{(a,b)}}{4R^2} - \sigma^{(a,b)}\,\epsilon \right] c_{k}^{(a,b)} \right\},
\end{multline}
with the starting values
\begin{align}
	c_{0}^{(a,b)} & = 1, & c_{1}^{(a,b)} = \frac{1}{2}\left(1+\frac{b-a}{2R}\right).
\end{align}
and 
\begin{align}
	\sigma^{(0,0)} & = 1,
	\\  
	\sigma^{(1,0)} = \sigma^{(0,1)} & = 1 + 5/(4-16R^2\epsilon), 
	\\
	\sigma^{(1,1)} & = 1 + 3/(1-4R^2\epsilon).
\end{align}
The $(0,0)$ and $(1,0)$ families contain ground-state and excited-state wave functions with an odd number of nodes, while the $(0,1)$ and $(1,1)$ families (which have a node at $u = 2R$) yield excited-state wave functions with an even number of nodes.

%%% TABLE II %%%
\begin{table*}
\caption{
\label{tab:quasi}
Closed-form solutions for the ground and first-excited states of two electrons on a ring.
}
\begin{ruledtabular}
\begin{tabular}{cccccc}
$j$	&	$n$	&	$(a,b)$			&	$P_{j,n}^{(a,b)}(u)$		&	$R_{j,n}^{(a,b)}$			&	$\epsilon_{j,n}^{(a,b)}$	\\	
\hline
1	&	0	&	$(1,0)$			&	$1$					&	$1/2$					&	$9/4$				\\
1	&	1	&	$(0,0)$			&	$1+u/2$				&	$\sqrt{3/2}$				&	$2/3$				\\
1	&	1	&	$(1,0)$			&	$1+(15-\sqrt{33})u/24$	&	$\sqrt{3(7+\sqrt{33})/8}$		&	$25(7-\sqrt{33})/96$		\\
1	&	2	&	$(0,0)$			&	$1+u/2+5u^2/92$		&	$\sqrt{23/2}$				&	$9/46$				\\
2	&	1	&	$(0,1)$			&	$1+(15+\sqrt{33})u/24$	&	$\sqrt{3(7-\sqrt{33})/8}$		&	$25(7+\sqrt{33})/96$		\\
2	&	1	&	$(1,1)$			&	$1+u/2$				&	$\sqrt{5/2}$				&	$4/5$				\\
2	&	2	&	$(1,1)$			&	$1+u/2+7u^2/132$		&	$\sqrt{33/2}$				&	$5/22$				\\
\end{tabular}
\end{ruledtabular}
\end{table*}

%%% FIGURE 1 %%%
\begin{figure}
	\includegraphics[width=0.4\textwidth]{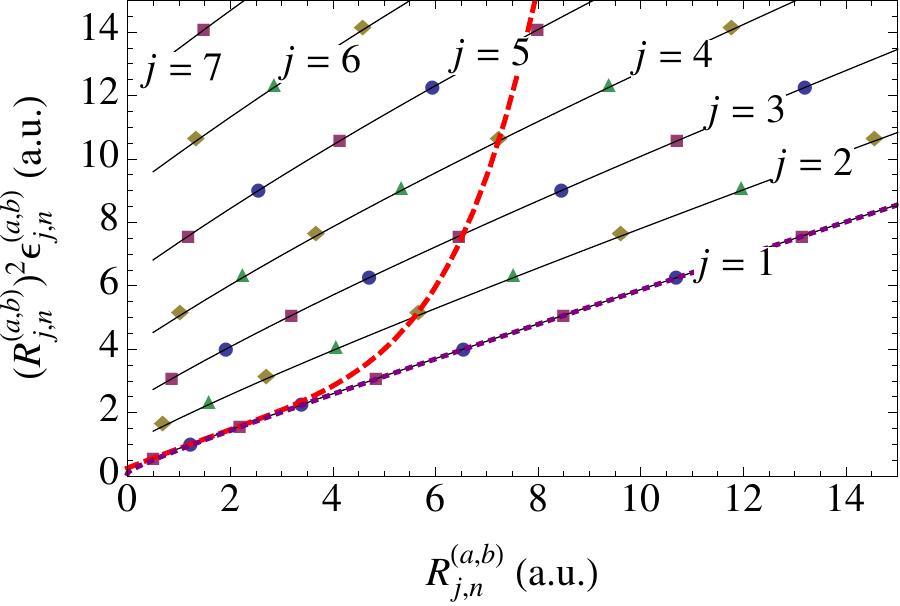}
	\caption{(Color online) Energies $\epsilon_{j,n}^{(a,b)}$ of the lowest $J=0$ states of two-electron quantum rings as a function of the radius $R_{j,n}^{(a,b)}$. Closed-form solutions in the $(0,0)$, $(1,0)$, $(0,1)$ and $(1,1)$ families are shown by blue dots, red squares, yellow diamonds and green triangles, respectively. The small-$R$ approximation Eq.~\eqref{E-smallR} (red dashed line) and large-R approximation Eq.~\eqref{E-largeR} (purple dotted line) are also shown.}
	\label{fig:quasi}
\end{figure}

%%% FIGURE 2 %%%
\begin{figure*}
	\subfigure[$j=1$ and $(a,b)=(0,0)$]{
	\includegraphics[width=0.2\textwidth]{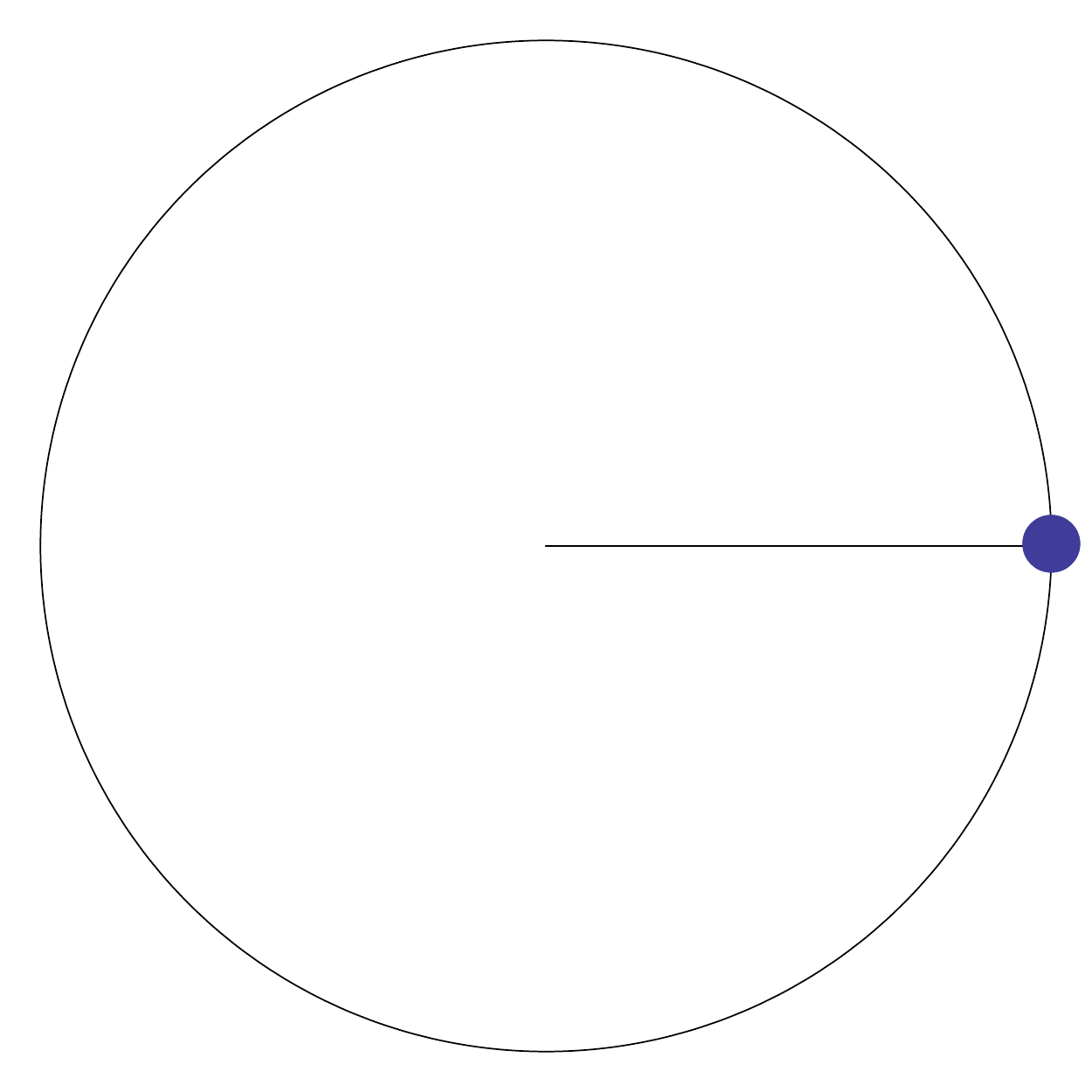}}
	\subfigure[$j=1$ and $(a,b)=(1,0)$]{
	\includegraphics[width=0.2\textwidth]{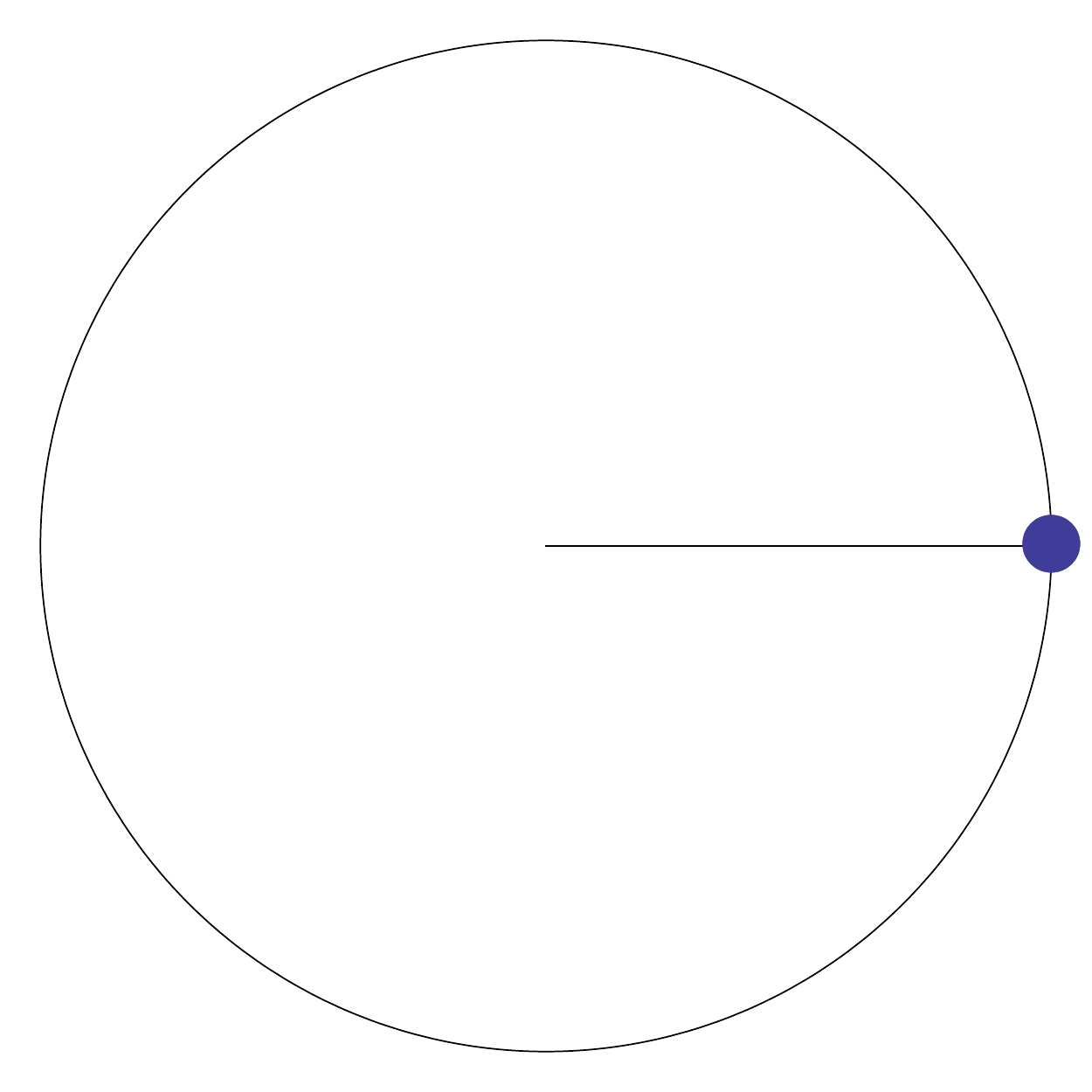}}
	\subfigure[$j=2$ and $(a,b)=(0,1)$]{
	\includegraphics[width=0.2\textwidth]{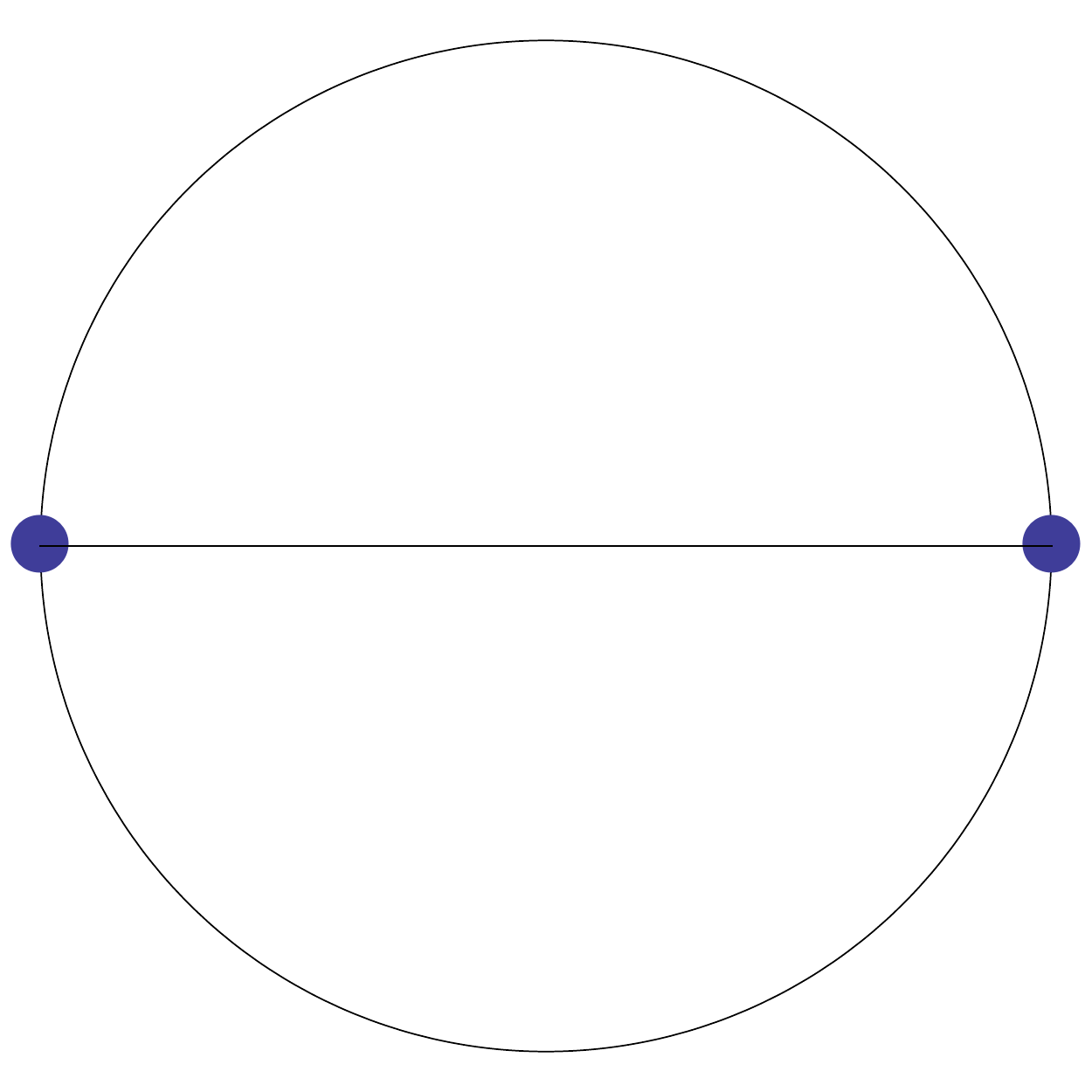}}
	\subfigure[$j=2$ and $(a,b)=(1,1)$]{
	\includegraphics[width=0.2\textwidth]{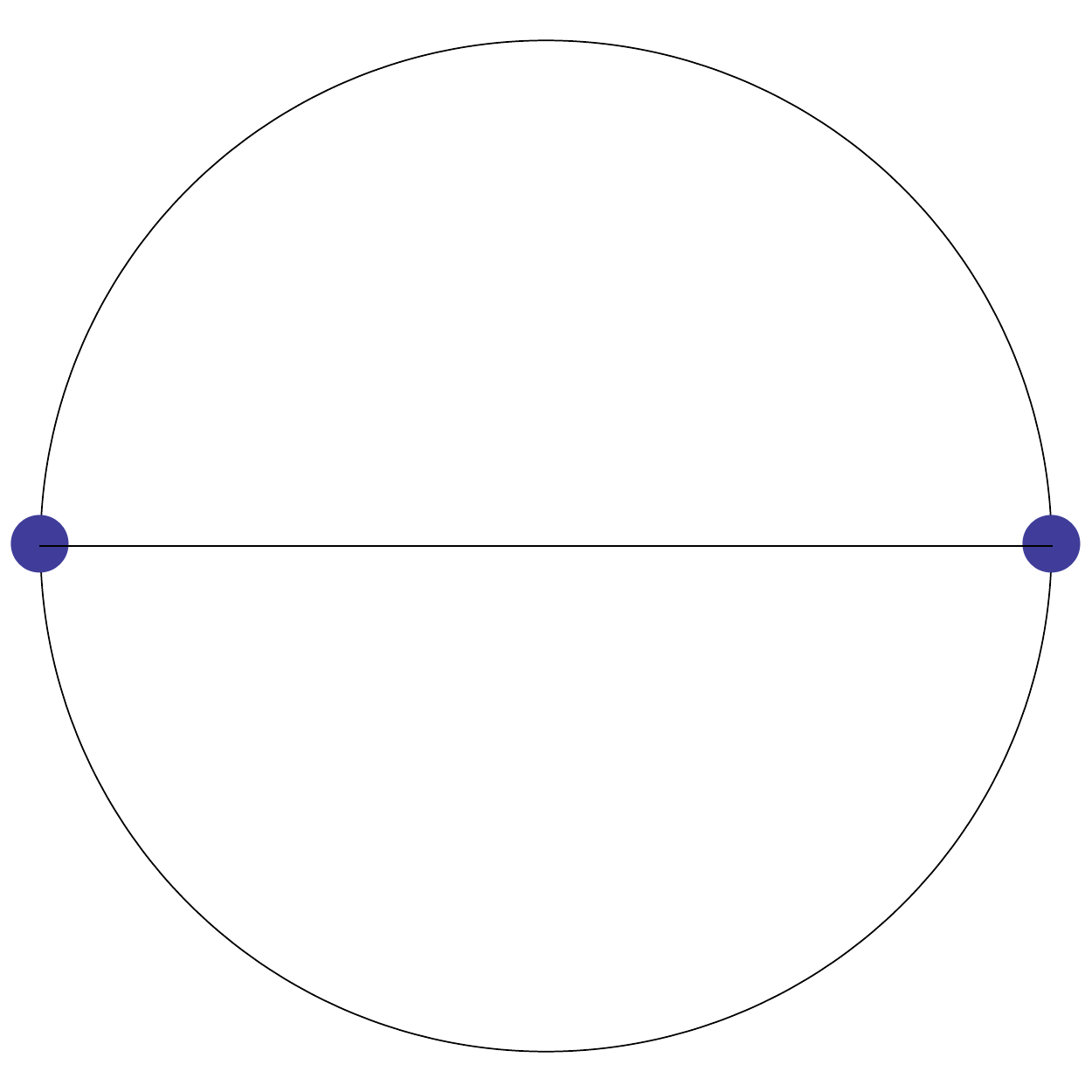}}
	\\
	\subfigure[$j=3$ and $(a,b)=(0,0)$]{
	\includegraphics[width=0.2\textwidth]{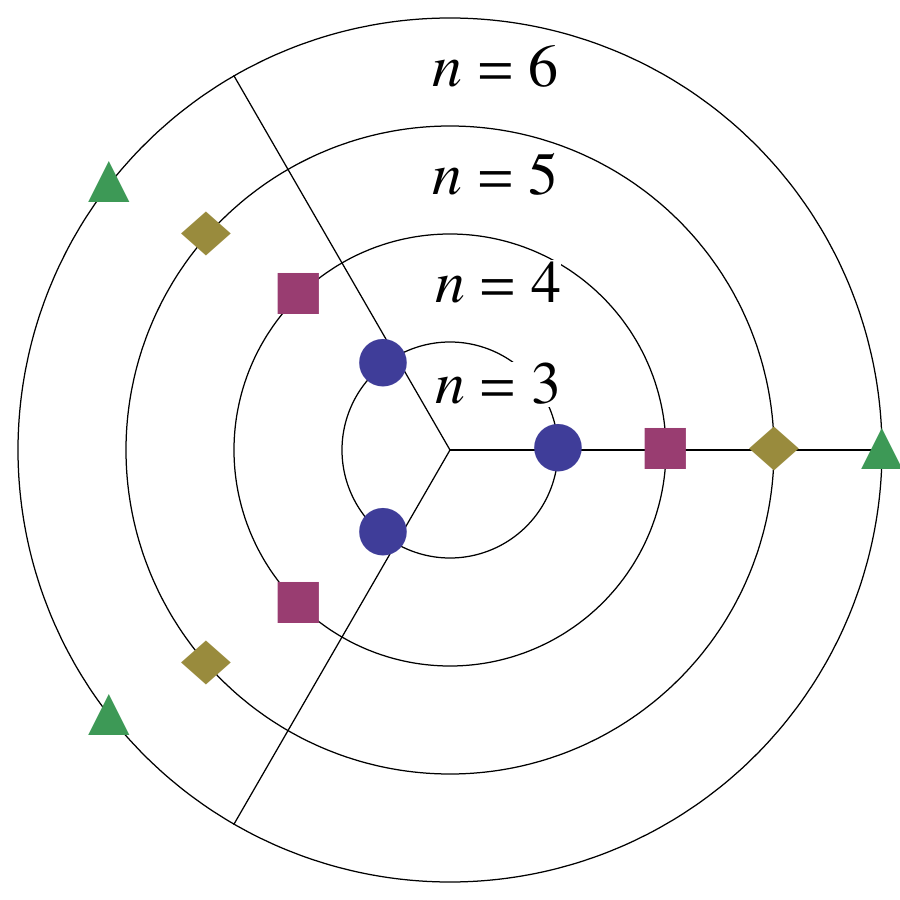}}
	\subfigure[$j=3$ and $(a,b)=(1,0)$]{
	\includegraphics[width=0.2\textwidth]{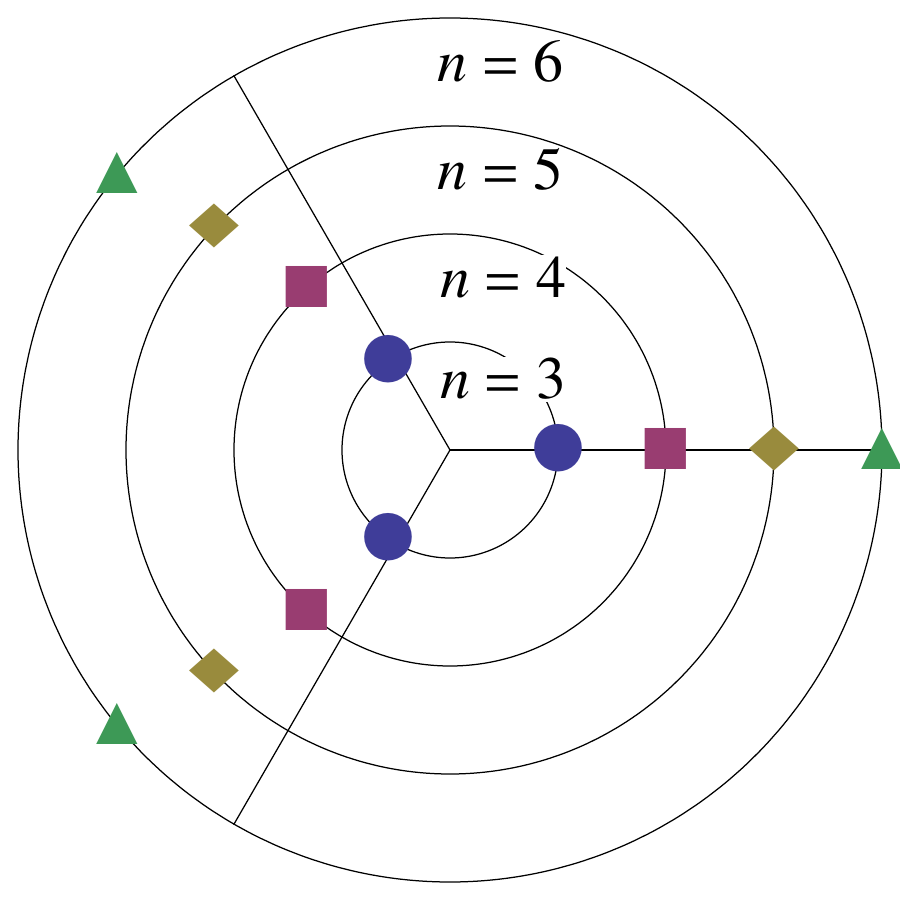}}
	\subfigure[$j=4$ and $(a,b)=(0,1)$]{
	\includegraphics[width=0.2\textwidth]{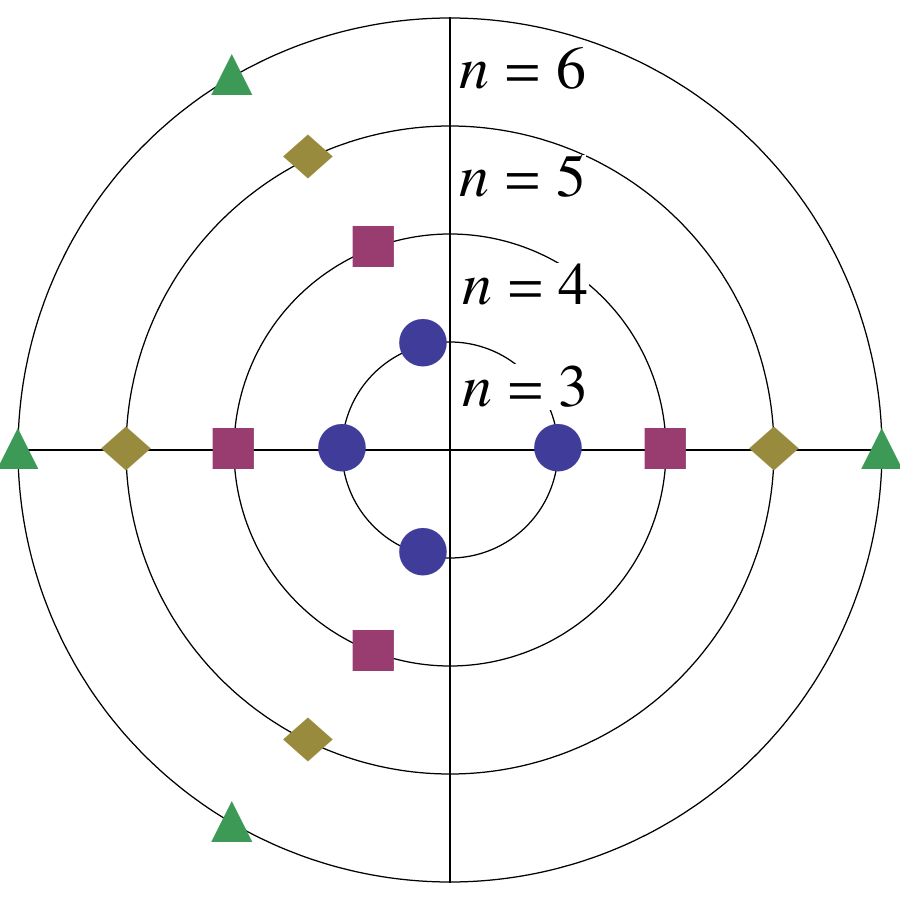}}
	\subfigure[$j=4$ and $(a,b)=(1,1)$]{
	\includegraphics[width=0.2\textwidth]{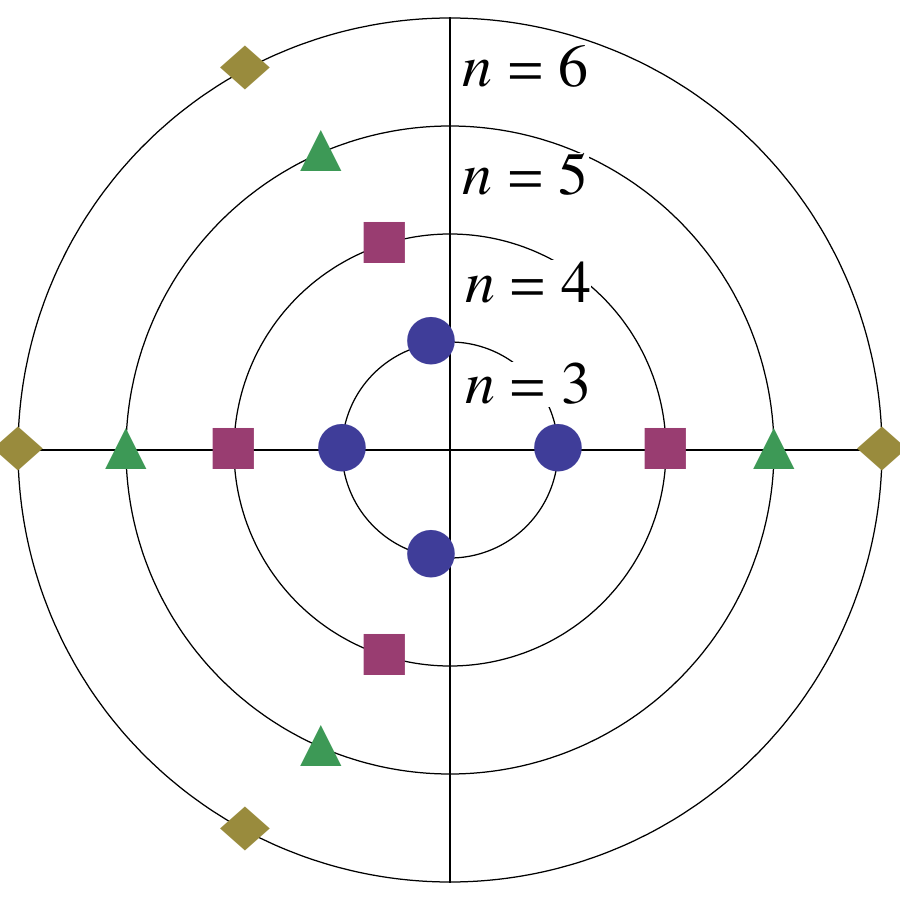}}
	\\
	\subfigure[$j=5$ and $(a,b)=(0,0)$]{	
	\includegraphics[width=0.2\textwidth]{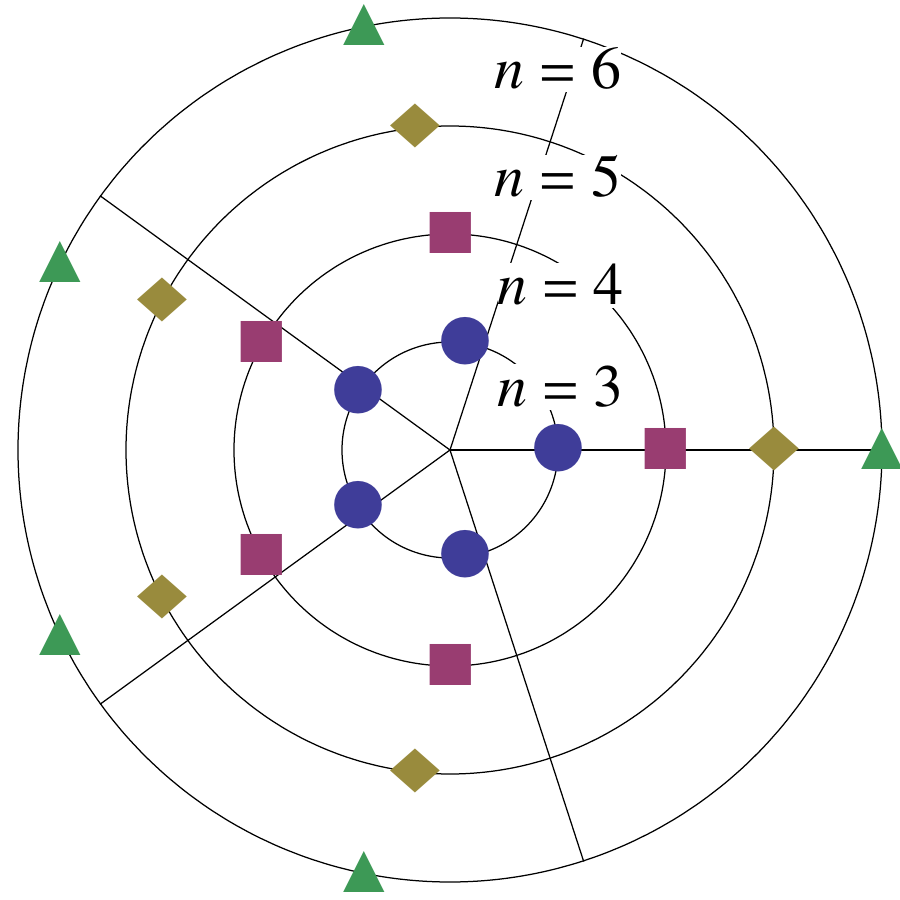}}
	\subfigure[$j=5$ and $(a,b)=(1,0)$]{
	\includegraphics[width=0.2\textwidth]{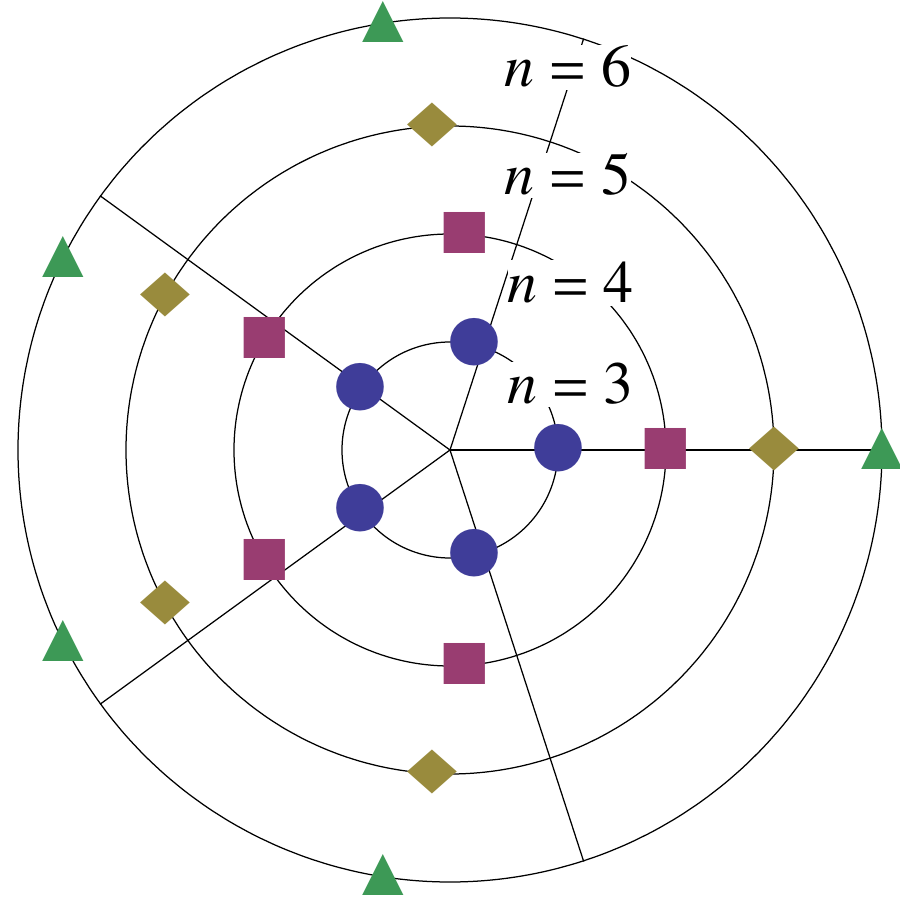}}
	\subfigure[$j=6$ and $(a,b)=(0,1)$	]{
	\includegraphics[width=0.2\textwidth]{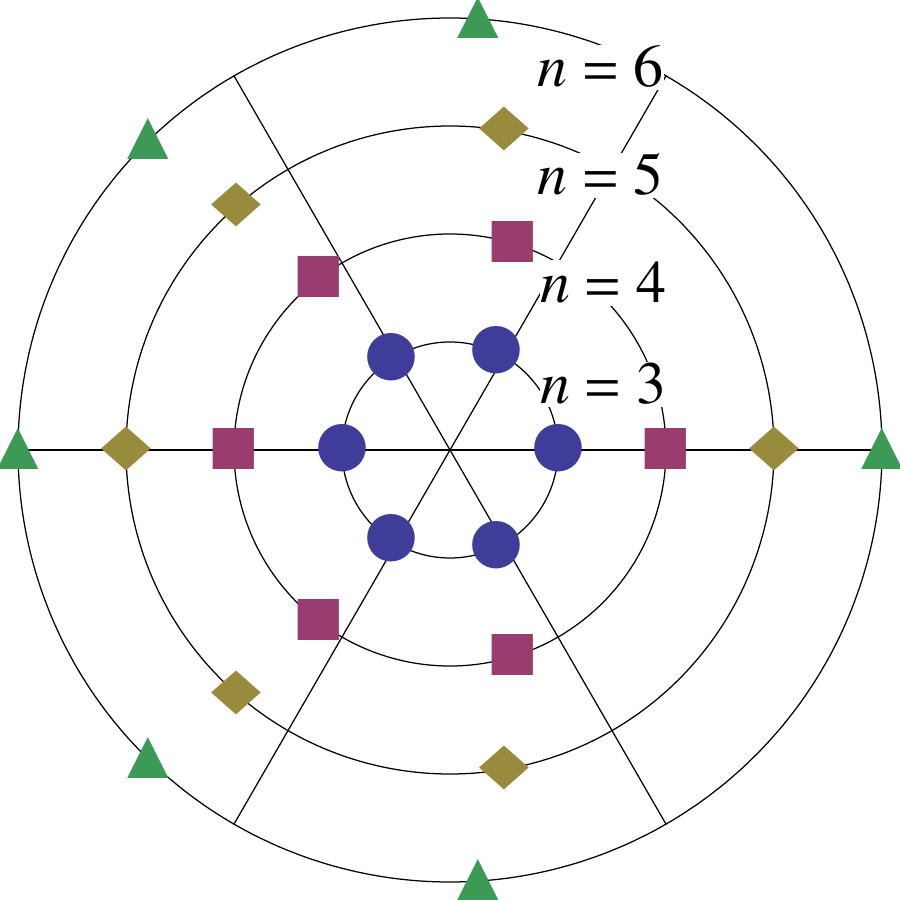}}
	\subfigure[$j=6$ and $(a,b)=(1,1)$]{
	\includegraphics[width=0.2\textwidth]{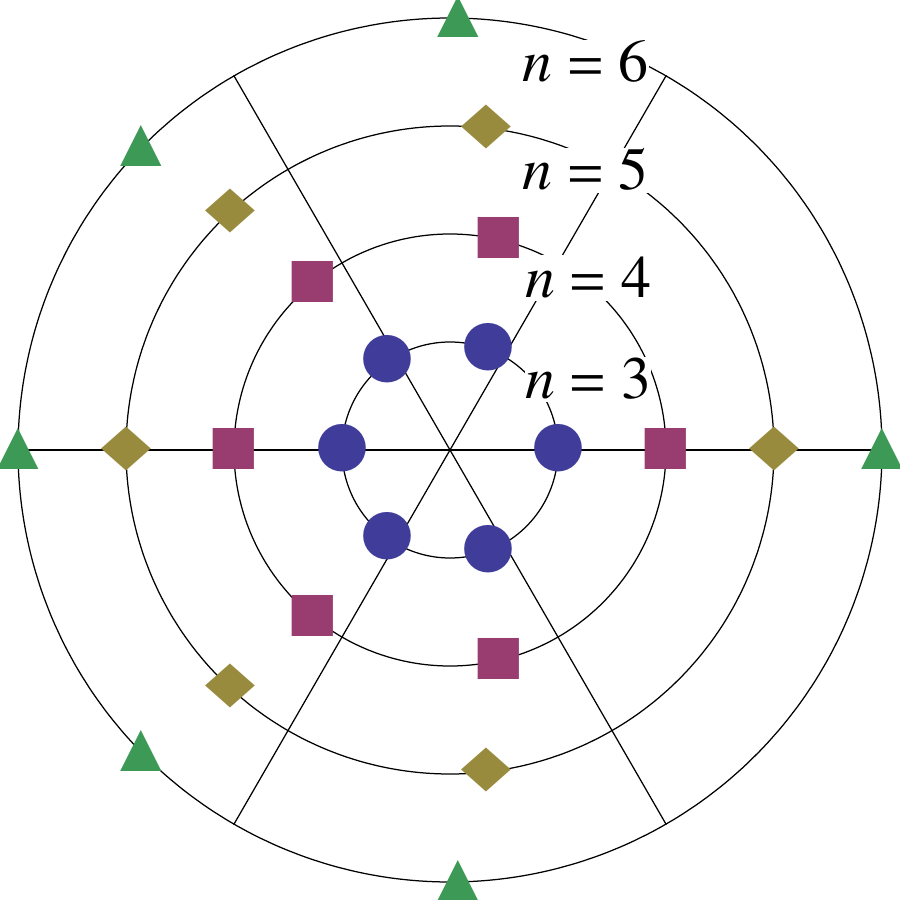}}
	\caption{(Color online) Node positions of closed-form solutions for the ground and excited states of two electrons on a ring. \alert{The black lines indicate the position of the non-interacting nodes.}}
	\label{fig:node}
\end{figure*}

%%% CLOSED-FORM SOLUTIONS %%%
{\em Closed-form solutions.---}For particular values of $R$, closed-form  solutions can be found.  The series \eqref{Pu} reduces to the $n$th-degree polynomial
\begin{equation}
	P_{j,n}^{(a,b)}(u) = \sum_{k=0}^{n} c_{k,j}^{(a,b)}\,u^k,
\end{equation}
if, and only if, $c_{n+1}^{(a,b)} = c_{n+2}^{(a,b)} = 0$.  The energy $\epsilon_{j,n}^{(a,b)}$ is a root of the polynomial equation $c_{n+1}^{(a,b)} = 0$ and the corresponding radius $R_{j,n}^{(a,b)}$ is found from \eqref{recurrence} to satisfy
\begin{equation} 
\label{E_S}
	4\,\sigma^{(a,b)}\,\epsilon_{j,n}^{a,b}\,\left(R_{j,n}^{(a,b)}\right)^2 =n(n+2+a+b) + \sigma^{(a,b)}.
\end{equation}
$\Psi_{j,n}^{(a,b)}$ is an exact intracular wave function with $j$ nodes.

Each $(a,b)$-family contains an infinite number of solutions, associated with distinct values of $R$ (Fig.~\ref{fig:quasi}). Both ground state and excited state wave functions can be obtained, and they are easily characterized by the number $j$ of nodes (Table \ref{tab:quasi}). The $(0,0)$ family contains polynomial solutions, while the three other families contain irrational solutions.  All behave as in \eqref{Kato} for small $u$.

The nodal patterns of the ground state (\textit{i.e.}~single node at $u=0$) and first excited state (nodes at $u=0$ and $2R$) are trivial \cite{Mitas06}, but those of the higher excited states are more complicated, and depend on the value of $R$ (Fig.~\ref{fig:node}). 

Taking \eqref{Phi-HF} as a zeroth-order wave function, one can use standard perturbation theory methods \cite{TEOAS09} to show that the small-$R$ (weak correlation) expansion of the ground-state ($j=1$) energy is 
\begin{multline}
\label{E-smallR}
	\epsilon = \epsilon_{\text{HF}}-0.026424+0.007241\,R
	\\
	-0.001966\,R^2+0.000492\,R^3+\ldots,
\end{multline}
which gives good agreement with the first four values of $R_{j,n}^{(a,b)}$ (Fig.~\ref{fig:quasi}).  At the other extreme, the large-$R$ (strong correlation) expansion is \cite{Loos10}
\begin{equation}
\label{E-largeR}
	\epsilon = \frac{1}{2R} + \frac{1}{4R^{3/2}} + \frac{5}{64R^2} + \ldots.
\end{equation}
Figure \ref{fig:quasi} reveals that Eq.~\eqref{E-largeR} is accurate over a much wider range of $R$ values than Eq.~\eqref{E-smallR}, as one might expect in such a strongly correlated one-dimensional system.

%%% CONCLUSION %%%
{\em Conclusion.---}In this Letter, we have shown that the Schr\"odinger equation for two electrons on a ring is solvable in closed form for a countably infinite number of $R$ values. We have demonstrated that, for each value of the angular momentum $J$, one is able to obtain polynomial and irrational solutions for both the singlet and triplet manifolds.  The latter are degenerate but exhibit different geometric (Berry) phase behavior. \alert{Although, we are not aware of any physical significance for these special values of the radius, they yield exact wave functions in both the weakly and strongly correlated regimes. This makes the present system particularly valuable for testing approximate methods in different correlation regimes. Following the approach developed in this Letter, analytical solutions for other interaction potentials can also be found.}  

We thank Joshua Hollett for fruitful discussions, and the Australian Research Council (Grants DP0984806, DP1094170 and DP120104740) for funding.

\end{document}